\documentstyle[11pt,newpasp,twoside]{article}
\def\LCDM{$\Lambda$CDM}
\def\eg{{\it e.g.}}
\def\etal{{\it et al.}}
\def\etc{{\it etc.}}

\def\edcomment#1{\iffalse\marginpar{\raggedright\sl#1\/}\else\relax\fi}
\reversemarginpar

\begin{document}
\title{What is the Evidence for Dark Matter?}
\author{J. A. Sellwood}
\affil{Department of Physics \& Astronomy, Rutgers University \\ 136 
Frelinghuysen Road, Piscataway, NJ 08855}

\begin{abstract}
Newtonian mechanics indicates that galaxies and galaxy clusters are
much more massive than we would have guessed from their luminosities,
with the discrepancy being generally attributed to dark matter halos.
An alternative hypothesis is that accelerations in very weak
gravitational fields are larger than predicted by Newton's laws, and
there is no need for dark matter.  Even though we do not currently
have a satisfactory theory associated with this rival hypothesis, we
can ask whether any observational tests could rule it out or prefer it
over the dark matter hypothesis.  Current evidence suggests that
neither hypothesis enjoys a decisive advantage over the other.  If
dark matter turns out to be the correct interpretation however, then
theories of galaxy formation face some quite severe fine-tuning
problems.
\end{abstract}

\section{Introduction}
The succcess of the currently favored \LCDM\ model for structure
formation (\eg\ Bahcall \etal\ 1999; White, this meeting) has
persuaded many people to abandon alternatives (\eg\ Binney, these
proceedings).  But the absence of a well-worked alternative should
never be sufficient reason to decide that a particular model is
correct.  Despite dark matter being a central ingredient of this
popular model, our only evidence for its existence is the
gravitationally inferred mass discrepancies, and it is becoming clear
that there are significant difficulties with its apparent properties
(\eg\ Ostriker \& Steinhardt 2003).  Furthermore, alternative gravity
ideas continue to meet with some success.  I therefore prefer to keep
an open mind until experimental evidence for one or other
interpretation of mass discrepancies becomes compelling.

Here I review the rationale for alternative gravity theories (\S2) and
argue (\S3) that galaxy formation theory does not provide decisive
evidence in favor of dark matter.  I then try to outline possible
experimental ways in which the two hypotheses to account for mass
discrepancies could be distinguished.  If dark matter is an elementary
particle, then it may be possible to detect it directly (see \S4).
Alternatively, two types of astronomical observations might favor one
or other interpretation: (a) the acoustic oscillations in the cosmic
microwave background (\S5), and (b) whether light is a good tracer of
the mass.  The second is a many faceted question, ranging from the
issue of whether the rotation curve of a galaxy can be predicted from
its luminosity profile (\S6), to whether the dark matter halos have a
different shape, or are misaligned with the light, or have
substructure not associated with luminous features (\S7).  Finally, I
review some theoretical arguments related to mergers (\S8).

\section{Modifications to basic laws}
The puzzles presented by dark matter and especially by dark energy are
severe, but not yet sufficient to compel us to reject the laws that
seem to require their existence.  Yet it is legitimate to explore
whether alternative laws could lead to a simpler and more beautiful
picture.  A simplified picture would be attractive only if it were
consistent with all current data, and compelling if it could make
testable predictions that differ from those of the standard paradigm.
Unfortunately, no suggestion as yet comes close to meeting these
ideals.

The apparent flatness of the universe (\eg\ Hinshaw \etal\ 2003) is
generally interpreted to imply a critical energy density in a
Friedmann model.  The current \LCDM\ model supposes the energy density
to be composed of an ugly mixture of $\sim70\%$ dark energy,
$\sim25\%$ dark matter, $\sim4\%$ baryonic matter, and maybe as much
as $1-2\%$ in neutrinos, although this last number is an upper limit
that may well be much less.  Not only is it unattractive to have
several components having similar densities, but the interpretation of
the dark energy as Einstein's cosmological constant requires us to
live at a special time just when the matter and dark energy densities
are comparable; matter was dominant until relatively recently, and
dark energy is now becoming dominant as the expansion reaccelerates.
Finally, the value of the vacuum energy density is lower than
na{\"\i}ve estimates based on elementary quantum mode counting by some
120 orders of magnitude, a discrepancy for which physics is completely
unable to account (\eg\ Weinberg 2000).  Tracking quintessence models
(\eg\ Wang \etal\ 2000) and cyclic universe models (\eg\ Khoury \etal\
2003) answer some of these objections, while Padmanabhan \& Choudhury
(2002) suggest that perhaps a single scalar field could account for
both the dark energy and the dark matter -- an idea that would seem to
merit further investigation.

The particularly severe problems presented by dark energy have driven
even mainstream physicists (\eg\ Arkani-Hamed \etal\ 2002; Carroll
\etal\ 2003) to speculate that it may instead reflect a breakdown of
general relativity on the horizon scale.  But there are far fewer
physicists who would argue that dark matter, for which the evidence is
also exclusively gravitational, is another manifestation of new
gravitational physics.  One reason for this different attitude is that
a weakly-interacting, massive particle (WIMP), of the kind predicted
by supersymmetry, would survive as a relic from the early universe
with an expected abundance of order the critical density (\eg\ Kolb \&
Turner 1990, \S5.6).

It has therefore been left to a rather renegade group of mostly
astronomers to argue for alternatives to dark matter, and several such
ideas have been reviewed at this meeting by Aguirre (these
proceedings).  By far the best known of these alternative ideas is
MOND, first proposed by Milgrom (1983), with a more sophisticated
formulation being given by Bekenstein \& Milgrom (1984).  Their
original proposal was a modification of Newton's law of gravity that
would automatically ensure asymptotically flat rotation curves for
galaxies and the Tully-Fisher law of the form $M \propto V_{\rm
flat}^4$, with no requirement for dark matter.  This proposal has led
to a fitting formula for galaxy rotation curves of remarkable power
(see below), but as a theory of gravity (or of inertia) it is very
poorly founded, probably wrong in details (at least), makes no clear
predictions about the gravitational deflection of light, and does not
remove the need for all dark matter in galaxy clusters (Aguirre,
Schaye \& Quataert 2001; Sanders 2003).

Partly for these reasons, I do not focus on MOND, but attempt to
enumerate possible tests that could distinguish between dark matter
and any generic alternative gravity hypothesis.

\section{Structure and galaxy formation}
The hierarchical clustering of galaxies is believed to have arisen
from almost homogeneous initial conditions through gravitational
instability.  It is well-known that Newtonian gravity arising from
baryonic perturbations alone could not drive the evolution from the
observed tiny fluctuations in the cosmic microwave background to the
highly non-linear clustering hierarchy of galaxies, groups, clusters
and voids we measure today; dark matter drives the clustering in the
standard picture.  However, Sanders (2001) argues heuristically, and
Nusser (2002) and Knebe \& Gibson (2003) show, that
stronger-than-Newtonian accelerations of the kind contemplated here do
accelerate the rate of evolution to yield an approximation to the
observed clustering hierarchy.  As yet, it is unclear whether the
results can be consistent with all data.  The weakness of this
argument is that we lack an appealing, generally covariant,
alternative theory of gravity on which to base a self-consistent
calculation without dark matter.  It is therefore not possible to use
structure formation arguments to distinguish dark matter models from
alternative gravity models, because we do not yet know how the
predictions in the two scenarios might differ.

While large-scale structure is a clear success for \LCDM, the
properties of galaxies on smaller scales are more problematic.
Sellwood \& Kosowsky (1999) give a now rather dated list of the
difficulties confronting galaxy formation in the standard \LCDM\
model.  Since other speakers at this meeting have expanded on these
difficulties, I do not try to update this list here.  Furthermore, it
is generally argued that the shortcomings of the current model do not
neccessarily imply its failure, but may instead reflect the fact that
the predictions are simply too crude.  It is clearly desirable to
continue to refine the predictions from the \LCDM\ model, but it seems
unlikely that all problems will evaporate as calculations improve.

\section{Direct detection of WIMPs}
WIMPs are the most popular dark matter candidate.  Their interaction
cross-section with normal baryonic matter, while extremely small, is
expected to be non-zero.  We may therefore be able to detect them
directly and a number of experiments are underway.  Aside from the
puzzling seasonal variation in their detection rate reported by the
DAMA collaboration (Bernabei \etal\ 2003) all other experiments have
so far failed to show a positive signal (\eg\ Akerib \etal\ 2003;
Sanglard \etal\ 2003; \etc), and some almost completely exclude a
particle that could have been responsible for the DAMA result.

The absence of a positive detection is of no great significance at the
current sensitivity levels.  Most experiments barely reach the
sensitivity required to detect even the most optimistic models for the
properties of WIMPs (\eg\ Cline 2003).  Sensitivity is being improved
all the time, but there is a long way to go before non-detections
would begin to rule out interesting WIMP candidates.

It is expected that WIMPs and their anti-particles should be present
in almost equal numbers.  Annihilation radiation will be emitted at an
enhanced rate from high-density regions, such as a dark matter cusp in
the center of a galaxy (Bertone \etal\ 2001).  Mayer-Hasselwander
\etal\ (1998) report an EGRET detection of an unidentified gamma ray
source in the Galactic center, but it is now believed not to be
coincident with the Galactic center direction and is too concentrated
to arise from annihilation radiation (Hooper \& Dingus 2002).  The
non-detection by EGRET does not tell us much about WIMPs in the Milky
Way halo, since the densest part of the predicted cusp could easily
have been erased by normal astrophysical processes (\eg\ Merritt
\etal\ 2002).  The increased sensitivity of the future GLAST satellite
may be sufficient to detect the expected diffuse gamma radiation (\eg\
Wai 2002; but see also Evans \etal\ 2003).  [Boehm \etal\ (2003)
speculate that the INTEGRAL detection of $511\;$keV radiation from the
Galactic Center could be annihilation radiation from a low-mass dark
matter particle.]  A convincing detection of diffuse annihilation
radiation would be conclusive evidence for the existence of dark
matter in the form of WIMPs.

\section{Acoustic oscillations in the CMB}
Temperature fluctuations in the cosmic microwave background are
believed to arise from accoustic oscillations seeded by primordial
density fluctuations.  Analysis of the power spectrum yields much
information about the nature of the universe (Jungman \etal\ 1996).
The first peak arises from the largest scale density variations that
have just had time to collapse since they entered the horizon.  The
second peak, at higher $l$, occurs for smaller scale waves that
entered the horizon at an earlier time and have had time to rebound to
the maximum rarefaction.  The third peak is on the scale of those
waves that have achieved a second collapse, \etc

Since baryonic oscillations are damped (Silk 1968), the heights of the
successive peaks should decrease monotonically in the absence of dark
matter.  Density variations in the dark matter, however, do not
oscillate because there is no pressure term -- the material interacts
with neither the photons nor the baryons.  The self-gravity of the
photon-baryon plasma oscillating in the potential variations caused by
the dominant dark matter causes the odd-numbered peaks (from
collapses) to be higher relative to the even ones (from rarefactions),
which is a generic signal of dark-matter driven oscillations (Hu \&
Sugiyama 1996, Hu \& White 1996).  Thus, we need to examine whether
the third peak is higher than the second.

Unfortunately, the height of the third peak is not yet strongly
constrained by the data.  The best fit model (Spergel \etal\ 2003)
places it at about the same height as the second peak, but the error
bars on the first year WMAP data are quite large at this scale.  The
unconventional analysis by \"Odman (2003) suggests that other
experiments may have systematic errors, since she suggests that the
relative height of the third peak depends on the frequency of the
observation, which should not be the case.  The relative heights of
the second and third peaks will not be determined until we have better
data -- perhaps soon to come from WMAP.  A third peak clearly higher
than the second would provide stronger, less model-dependent, evidence
for dark matter than is currently indicated by the first year WMAP
data (Spergel \etal\ 2003; see also McGaugh 2003).

\section{Light as a predictor of mass}
There is no doubt that Newtonian mechanics requires mass-to-light
ratios that increase with increasing scale (\eg\ Bahcall \etal\ 1995).
Modified gravity theories, which are proposed in order to account for
this fact, actually require much more: a unique relationship
between the baryonic mass and the dynamical mass.  This stringent
requirement leads to a number of specific predictions that makes them
much more easily falsified than is the dark matter hypothesis.

The relationship between the observed baryonic mass and the inferred
Newtonian dynamical mass can, in principle, depend on spatial scale,
or acceleration (as in MOND), or some other property, but the relation
must be universal.  Variations in the baryonic/dark mass ratio in
similar objects could arise in dark matter models, but cannot be
tolerated in modified gravity models.  Two well-established examples
where this holds are the Tully-Fisher relation for galaxies and the
absence of bias in the galaxy distribution; the 2dF galaxy redshift
survey team (Verde \etal\ 2002) conclude that ``optically selected
galaxies do indeed trace the underlying mass distribution.''
Estimates of bias on smaller scales (\eg\ Pen \etal\ 2003) are still
subject to significant uncertainties.

Furthermore, modified gravity theories require the gravitational
potential well to reflect the shape and orientation of the luminous
matter.  Dark matter halos, on the other hand, can have much more
general shapes and may also be misaligned with the orientation of the
luminous matter.  These predictions are confronted by available data
in this section.

\subsection{Detailed rotation curve shapes}
Kalnajs's (1983) dramatic demonstration that the inner parts of galaxy
rotation curves can be predicted from the light distribution by a
simple mass-traces-light model has been amply confirmed in many
subsequent investigations (Kent 1986; Buchhorn 1992; Broeils \&
Courteau 1997; Palunas \& Williams 2000; Sancisi, these proceedings).
However, such models based on Newtonian dynamics generally fail at
larger radii, where the predicted rotation curve falls below that
observed.

The outstanding triumph of MOND is that it has yielded a formula that
is able to predict the entire detailed rotation curve of a disk galaxy
from the light profile.  This achievement goes considerably further
than its original motivation and works best for the galaxies for which
the best data are available, as first shown for a sample of 10
galaxies by Begeman, Broeils \& Sanders (1996).  Sanders \& McGaugh
(2002) increase the sample to $\sim75$ well-studied disk galaxies.

MOND rotation curve fits have only the mass-to-light ratio,
$\Upsilon$, for the stellar component(s) as a free parameter.  In some
cases, the distance to the galaxy might require adjustment because the
actual acceleration within the galaxy, $V^2/R$, is distance-dependent,
while the characteristic acceleration $a_0$ that enters the fitting
formula is not.  Generally the fits are quite successful for the
standard distance to the galaxy, but adjustments to improve the fits
can be as large as a factor 2 in a few cases.  A distance required for
a good MOND fit could, in principle, be excluded by independent
distance indicators, and the few cases where this could potentially
invalidate MOND are discussed in Bottema \etal\ (2002).

Milgrom \& Sanders (2003) demonstrate that the declining rotation
curves in elliptical galaxies recently announced by Romanowsky \etal\
(2003) are also in accord with MOND predictions.  Gerhard \etal\
(2001) argue that mass discrepancies begin to appear in elliptical
galaxies at much stronger accelerations, $\sim 10 a_0$, although their
conclusion is also strongly challenged by Milgrom \& Sanders (2003).
If it can be convincingly demonstrated that the value of $a_0$ is not
universal, the success of MOND for spiral galaxies could no longer be
interpreted as a manifestion of an unknown universal physical law.

It should be emphasized that conventional dark matter fits have three
free parameters: $\Upsilon$, and two halo parameters, such as the core
radius, $r_0$, and $V_{\rm flat}$ for the psuedo-isothermal halo, or
the concentration, $c$, and $V_{200}$ for an NFW halo.  The success of
MOND fits, with essentially only a single parameter, therefore cannot
be dismissed lightly.  If dark matter halos are, in fact, present in
these galaxies, then the impressive success of the MOND fitting
formula implies an intimate connection between the dark and luminous
matter that presents a daunting fine-tuning problem for dark matter
models.

Kaplinghat \& Turner (2002) attempt to show that a characteristic
acceleration of the MOND value emerges ``naturally'' from the CDM
model, but their claim is weak.  They show only that the acceleration
within a galaxy at which dark matter begins to dominate is of the
order of MOND's $a_0$.  Since their analysis uses the Gaussian initial
fluctuation spectrum, they inevitably predict a spread in $a_0$ that
is not required and they fail utterly to account for the spectacular
accuracy of the MOND fits to the detailed rotation curve shapes.

\subsection{Dark lenses?}
The case for dark matter would be essentially made if we were able to
discover objects with extremely high, almost infinite, M/L.  The only
way such objects might be discovered is through gravitational lensing.
Schneider (these proceedings) remarks that there are no known examples
of strong lensing by a dark obect, but a few such claims have been
made from weak lensing.

Probably the strongest claimed detection of a dark cluster is by Erben
\etal\ (2000), who find a ``robust'' lensing signal suggesting a
previously unidentified dark mass concentration of perhaps
$10^{14}\;\hbox{M}_\odot$.  There is no visible galaxy cluster and
little X-ray emission from the position of this apparent mass
concentration.  However, it is hard to imagine how a real dark
cluster-mass object could have avoided collecting enough hot gas to be
X-ray bright.  It is possible that the lensing signal is spurious and
arises from intrinsic alignments among the background galaxies.

Weak lensing surveys, \eg\ by Tyson's group (Jarvis \etal\ 2003),
should reveal more such cases if they are common, but none has shown
up so far.  In fact, the lensing mass does not have to be completely
dark to rule out alternative gravity theories, large variations in M/L
between similar lensing-mass clusters would be sufficient.

\section{Halo shapes}
Simulations of large-scale structure formation have revealed that the
dark matter halos of galaxies and clusters are expected to be
triaxial.  The detailed distribution of halo shapes reported by the
different groups varies slightly (see reviews by Springel and by Jing,
these proceedings), but most authors find many nearly prolate shapes.
It is also known (Dubinski 1994) that baryonic infall makes the halo
more nearly axisymmetric.

A na\"\i ve prediction from alternative gravity theories might be that
the potential is flattened near a flattened light distribution and
becomes more nearly spherical further away, since the higher order
moments of the potential drop off more quickly.  Aguirre (these
proceedings) points out that this simple-minded expectation is not
necessarily true, especially in MOND.

The meager data on the shapes of dark matter halos are reviewed by
Sackett and by Arnaboldi (these proceedings).  I continue to be
impressed by the absence of a clear detection of a strongly
non-axisymmetric extended HI disk that can be attributed to a triaxial
dark matter halo.  Bekki \& Freeman (2002) speculate that spirals in
the outer disk of NGC~2915 arise from a rotating triaxial halo, but
other intepretations are possible (\eg\ Masset \& Bureau 2003).  The
startlingly round outer ring of IC 2002 (Franx \etal\ 1994) remains an
isolated case.  The lack of other cases could, perhaps, be a selection
bias if gas in a non-axisymmetric halo is driven inwards by shocks
until it reaches a closely axisymmetric potential.

\subsection{Misalignments}
Kochanek (2002) could find no evidence for misalignments between the
luminous matter and the lensing mass distribution in a number of
strong lens systems, and concluded that ``Mass traces light.''
Perhaps yet more impressive is the claim by Hoekstra \etal\ (2003) of
a detectable flattening of the dark matter halos in a statistical
analysis of weak lensing data from stacking a large number of
galaxies.  If halos were mostly prolate, and the density axes at
larger radii were twisted at random angles from those in the inner
parts, the flattening found by Hoekstra \etal\ (2003) should not be
detectable.  (See Aguirre, these proceedings, for further discussion
of their paper.)

Buote \etal\ (2002), on the other hand, argue that alternative
gravity theories can be excluded by their claim of misalignment
between the flattened x-ray halo of NGC~720 and the optical image.  It
is noteworthy that their estimate of the shape and orientation of the
halo changed greatly from their earlier estimates based on ROSAT data
to this recent anaylsis of Chandra data, due mainly to the elimination
of point sources.  The lumpy nature of the x-ray isophotes in even
their cleaned Chandra image may indicate further contamination by
fainter point soucres, or it may indicate that the halo gas is subject
to non-gravitational forces, or has has yet to settle.  While this one
case is not sufficiently compellling to rule out alternative gravity
models, their research program is one of the few with this goal.

\subsection{Substructure}
Dalal \& Kochanek (2002) argue that the flux ratio anomalies observed
in a significant number of quadruple image lens systems indicate the
presence of substructure in the dark matter halo.  Convincing evidence
for the mini-halos predicted in CDM models would constitute decisive
support for dark matter.  Mao (these proceedings) reviews the
evidence, finding a number of puzzles, but concludes that mini-halos
are the most likely explanation for flux anomalies, while Schechter
(these proceedings) argues that alternative explanations are hard to
rule out.

\section{Galaxy mergers}
Binney \& Toomre asked whether modified gravity theories could be
consistent with galaxy mergers; specifically, whether galaxies that
fall together could merge if there are no DM halos to take up the
orbital energy and angular momentum?  Binney \& Tremaine (1987) argue
that a pair of similar, spherical galaxies approaching each other will
dissipate their orbital energy and angular momentum and merge within a
Hubble time for some limited range of specific orbital energies and
angular momenta.  In a modified gravity picture, the two galaxies
would accelerate towards each other more strongly, and would therefore
acquire much greater kinetic energy and angular momentum for their
masses, making a merger less likely.  The effect will scarcely be
compensated by increased orbit-to-internal exchanges during the
interaction, because forces during the encounter are closer to the
Newtonian regime.

Mergers can occur even without dark halos, for sufficiently slow
passages, as demonstrated for Newtonian dynamics by Toomre \& Toomre
(1972).  Modified gravity is unlikely to change this conclusion, but
will make such encounters rarer.  The question then becomes a
quantitative one of whether the merger rate will be sufficiently
suppressed by modified gravity to be inconsistent with that observed.
A clear answer to this question requires a detailed calculation that
has yet to be performed, and will depend on the nature of the
modification.  Two further points are worth making: First, Toomre's
(1977) idea that elliptical galaxies are products of mergers seems to
be holding up, but most were probably in place already at moderately
high redshift (\eg\ van Dokkum \& Ellis 2003), perhaps before the time
that departures from Newtonian dynamics become significant (Sanders
2001).  Second, Hickson compact galaxy groups seem to undergo rapid
mergers (Barnes 1989) in conventional dark matter models, making the
present-day frequency of these groups difficult to understand -- but
see also White (1990); modified gravity theories should ease this
difficulty.

Studies of individual cases are always revealing.  Dubinski \etal\
(1999) argue that tidal tails require rather low density dark matter
halos in conventional dynamics.  Comparable studies with modified
gravity have not been reported, but would seem to be warranted.  In
particular, it would be very interesting to test whether simulations
of merging systems with dark matter can account for the dramatically
more concentrated stellar component in elliptical galaxies reported by
Romanowsky \etal\ (2003).

\section{Conclusions}
This very brief survey of the available data has turned up no really
decisive evidence either in favor of dark matter or that could rule
out alternative gravity theories.  To form a scorecard, I regard the
strongest evidence in favor of dark matter as the CMB power spectrum
and the explanation of flux ratio anomalies in quadruple lenses in
terms of halo substructure.  On the other hand, new gravitational
physics is favored by the extraordinarily clear relationship between
the light and mass distributions in galaxies, and the lensing evidence
(Kochanek 2002, Hoekstra \etal\ 2003) that the mass distribution seems
to be aligned with the light -- Buote \etal\ (2002) notwithstanding.  The
evidence from halo shapes is too fragmentary to put in either column.

It needs to be stressed that the success of MOND fits to rotation
curves and the alignment of lensing mass with the visible mass
distribution create very serious fine-tuning problems if mass
discrepancies are indeed caused by dark matter halos.

As always, more data will help; it is especially desirable to pursue
M/L variations, misalignments, substructure, \etc, since alternative
gravity ideas can be more easily falsified by such evidence than can
dark matter models.  Decisive evidence for dark matter may come soon
from a firm measurement of the relative heights of the 2nd and 3rd
peaks in the CMB, possibly even from WMAP, or from a detection of
annihilation radiation by the future GLAST mission.  In the present
absence of decisive evidence on either side, the peculiarities
highlighted here seem significant enough to encourage further attempts
to develop alternative gravity theories from which testable
predictions could be made.

\acknowledgments
I thank Arthur Kosowsky and Moti Milgrom for comments on the
manuscript.  This work was supported by grants from NASA (NAG 5-10110)
and from NSF (AST-0098282).

\end{document}